# Enhancement of ferroelectric performance in PVDF:Fe$_3$O$_4$ nanocomposite based organic multiferroic tunnel junctions

Xue Gao,[1,2,3] Shiheng Liang,[1] Anthony Ferri,[4] Weichuan Huang,[5] Didier Rouxel,[1] Xavier Devaux,[1] Xiao-Guang Li,[5] Hongxin Yang,[6] Mairbek Chshiev,[7] Rachel Desfeux,[4] Antonio Da Costa,[4] Guichao Hu,[8] Mathieu Stoffel,[1] Abir Nachawaty,[1] Chunping Jiang,[2,3] Zhongming Zeng,[2,3] Jian-Ping Liu,[2,3] Hui Yang[2,3] and Yuan Lu[1]*

[1]*Université de Lorraine, CNRS, Institut Jean Lamour, UMR 7198, campus ARTEM, 2 Allée André Guinier, 54011 Nancy, France*
[2]*School of Nano Technology and Nano Bionics, University of Science and Technology of China, 96 Jinzhai Road, Baohe, Hefei 230026, P.R.China*
[3]*Suzhou Institute of Nano-Tech and Nano-Bionics, Chinese Academy of Sciences, 215123, Suzhou, P.R.China*
[4]*Univ. Artois, CNRS, Centrale Lille, ENSCL, Univ. Lille, UMR 8181, Unité de Catalyse et Chimie du Solide (UCCS), F-62300 Lens, France*
[5]*Hefei National Laboratory for Physical Sciences at Microscale, Department of Physics, University of Science and Technology of China, Hefei 230026, P.R.China*
[6]*Key Laboratory of Magnetic Materials and Devices, Ningbo Institute of Materials Technology and Engineering, Chinese Academy of Sciences, Ningbo 315201, P.R.China*
[7]*Univ. Grenoble Alpes, CEA, CNRS, Spintec, 38000 Grenoble, France*
[8]*School of Physics and Electronics, Shandong Normal University, Jinan 250358, P.R.China*

*\* Corresponding author: yuan.lu@univ-lorraine.fr*

**Abstract:**

We report on the fabrication of organic multiferroic tunnel junction (OMFTJ) based on an organic barrier of Poly(vinylidene fluoride) (PVDF):Fe$_3$O$_4$ nanocomposite. By adding Fe$_3$O$_4$ nanoparticles into the PVDF barrier, we found that the ferroelectric properties of the OMFTJ are considerably improved compared to that with pure PVDF barrier. It can lead to a tunneling electroresistance (TER) of about 450% at 10K and 100% at room temperature (RT), which is much higher than that of the pure PVDF based device (70% at 10K and 7% at RT). OMFTJs based on the PVDF:Fe$_3$O$_4$ nanocomposite could open new functionalities in smart multiferroic devices via the interplay of the magnetism of nanoparticle with the ferroelectricity of the organic barrier.

The use of a ferroelectric thin film as barrier in a magnetic tunnel junction (MTJ) adds an additional level of functionality. In such a multiferroic tunnel junction (MFTJ), the tunneling electroresistance (TER) and tunneling magnetoresistance (TMR) effects coexist,[1] which could possibly lead to new applications such as multi-level data storage devices.[2,3] Compared to inorganic spintronics, its organic counterpart is very appealing because of the long spin lifetime of charge carriers in addition to their relatively low cost, flexibility, and chemical diversity.[4,5,6] As one of the most typical ferroelectric polymers, poly (vinylidene fluoride) (PVDF) has been widely used in modern electronic systems and devices, whose ferroelectric performance relies mainly on its β phase content.[7] Moreover, PVDF and the related copolymers can form high quality ordered layers[8] and exhibit robust ferroelectricity down to monolayer thickness.[9] Their performances are comparable to that of perovskite oxide ferroelectrics,[10] making them suitable as tunneling barriers for organic multiferroic tunnel junctions (OMFTJs).[11,12] Recently, our group demonstrated the ferroelectric control of the "spinterface" spin-polarization in the PVDF or P(VDF-TrFE) based OMFTJs,[13,14] which opens new functionalities in controlling the injection of spin polarization into organic materials via the ferroelectric polarization of the barrier. Further improvement of the ferroelectric properties of the OMFTJ devices is urgently required to promote potential memristor and spintronics applications.

Magnetite $Fe_3O_4$ has recently gained an increasing interest since bulk $Fe_3O_4$ has a high Curie temperature ($T_c$~850K) and is almost fully spin-polarized at room temperature (RT). Both properties are of great interests for applications in giant magneto-electronic and spin-valve devices.[15] It has been reported previously that ferroferric oxide ($Fe_3O_4$) spherical shaped nanoparticles are excellent fillers for high-dielectric-constant polymer composites.[16,17,18] Some researchers found that the ferroelectric properties of PVDF are improved by adding $Fe_3O_4$ nanoparticles.[19,20,21,22] This is generally explained by the interaction between the nanoparticles and the $CH_2$ groups of the polymer chains, which promotes the nucleation of the polar β phase of PVDF.[23,24] However, up to now, there are still no reports on the impact of $Fe_3O_4$ nanoparticles on ferroelectric properties of PVDF based ferroelectric tunneling junction (FTJ) or OMFTJ.

In this Letter, we report on the fabrication of OMFTJs based on $La_{0.6}Sr_{0.4}MnO_3$/PVDF:$Fe_3O_4$ nanocomposites/Co structures. By dispersing $Fe_3O_4$ nanoparticles into the PVDF barrier, we found that the ferroelectric properties of PVDF are much improved, showing a smaller polarizing voltage for polarization switching. Consequently, the OMFTJs based on the nanocomposites demonstrate superior performances: much larger amplitude of the TER and the improved thermal stability compared to the OMFTJs with a pure PVDF barrier.

OMFTJs based on $La_{0.6}Sr_{0.4}MnO_3$/PVDF:$Fe_3O_4$ nanocomposites/Co/Au structures were prepared as follows. Firstly, a 50 nm thick $La_{0.6}Sr_{0.4}MnO_3$ (LSMO) film was grown epitaxially on a $SrTiO_3$ (STO) (001) substrate at 750ºC using DC magnetron sputtering. The LSMO layer was then etched by a hydrogen chloride solution (37%) to obtain a pattern consisting of 200 μm width bars as the bottom electrodes. Secondly, PVDF:$Fe_3O_4$ nanocomposite films were prepared by spin-coating with the following procedures. The PVDF solution was prepared by dissolving PVDF powder into a dimethylformamide (DMF) solution with a concentration of 20mg/mL. The $Fe_3O_4$ nanoparticles diluted in toluene (5nm nominal size purchased from Sigma-Aldrich) with a concentration of 5mg/mL were added into the PVDF solution with a ratio of 1:1 in volume. The mixture was sonicated (bath-type sonication) for 5h. The mixed suspension was then spin-coated onto the pre-patterned LSMO/STO substrate with a speed of 3000 RPM for 1 min. Subsequently, the as-coated film was annealed at 150ºC in Ar atmosphere for 2h to improve the crystallization of the ferroelectric β phase.

The thickness of the PVDF:Fe$_3$O$_4$ barrier was checked by the profilometer to be about 18 nm. Thirdly, the sample is then loaded into a molecular beam epitaxy (MBE) system (with a base pressure of 1×10$^{-10}$ Torr) for the growth of the top Co/Au electrodes. 10 nm thick Co and 10 nm thick Au were grown sequentially by e-beam evaporation with an *in-situ* shadow mask. To minimize the metal diffusion into organic materials,[25,26] the temperature of the substrate was maintained at around 90K during the growth of the top electrode. The final junction, that is schematically shown in Fig. 3(a), has a typical size of about 200×200 μm$^2$.

The surface morphologies of both pure PVDF and PVDF:Fe$_3$O$_4$ nanocomposite films were compared by atomic force microscopy (AFM) (Fig. 1(a) and 1(b), respectively). The pure PVDF surface exhibits a needle-like crystalline feature with a typical width of about 20 nm and a typical length of about 50 nm, which is a characteristic of the ferroelectric β phase of PVDF.[27] The root-mean square (RMS) roughness is about 3.2 nm for a 0.8×0.8 μm$^2$ scanning area. For the PVDF:Fe$_3$O$_4$ nanocomposite film, the surface roughness increases to reach a RMS of 7.1nm for a 0.8×0.8 μm$^2$ scanning area. One can observe spherical shaped Fe$_3$O$_4$ nanoparticle inclusions, which are randomly distributed in the PVDF layer. The size of the Fe$_3$O$_4$ inclusion particles is not homogenous, ranging from nm to μm size due to the aggregation of the magnetic nanoparticles. Conductive AFM (c-AFM) measurement is the most direct way to identify the existence of pinholes. Nanoscale conductivity variations were probed through current mapping experiments using the conductive-AFM technique. A Ti/Ir-coated silicon tip and cantilever with a stiffness of 3N·m$^{-1}$ were used. DC bias voltages ranging from -10 and +10V between the grounded AFM tip and the bottom electrode were applied during scanning. The electrical conductivity map recorded over 20×20 μm² large area are displayed in Fig. 1(c) and 1(d) respectively for the samples with pure PVDF and PVDF:Fe$_3$O$_4$ nanocomposite films. No conducting path is obtained for an applied bias of +5V, as demonstrated by the uniform contrast associated to insignificant current. Such absence of conduction signal was confirmed in several places of the sample surface regardless the applied bias voltage when varying between -10 and +10V. This confirms that the contribution of pinhole on transport is very limited despite the inhomogeneity of particle size. In addition, it also proves that the polarizing voltage around ±4V range does not induce a damage of organic barrier with the formation of pinholes and leakage currents.

The ferroelectric properties of both pure PVDF and PVDF:$Fe_3O_4$ nanocomposite films were compared by piezo-response force microscopy (PFM). In the PFM maps shown in Fig. 1(e) and 1(f), the films were firstly polarized with a positively biased tip over the area indicated with a dashed square. Then, a negative poling bias was subsequently applied in the middle area to reverse the polarization. As observed from the strong contrast of PFM phase in the two polarization states, the polarizations of both pure PVDF and PVDF:$Fe_3O_4$ nanocomposite films are quite homogenous. Moreover, a complete switching of the polarization can be achieved in both cases. Nevertheless, the difference in ferroelectric properties are highlighted by the local PFM phase and amplitude hysteresis loops, which are shown in Fig. 1(g) and 1(h). As observed in Fig. 1(h), the coercivity of the polarizing voltage is about 0.2 V, that is much smaller than the corresponding value (1.2V) measured on the pure PVDF film (Fig. 1(g)). This indicates that the switching of the polarization is much easier in PVDF:$Fe_3O_4$ nanocomposite films due to the improvement of their ferroelectric properties.

To investigate the magnetic properties of the $Fe_3O_4$ particles dispersed inside the PVDF, we have measured the magnetization of a Si/Au/PVDF:$Fe_3O_4$ sample with a superconducting quantum interference device (SQUID) magnetometer. Fig. 2 shows the temperature dependent magnetic hysteresis loops measured of the Si/Au/PVDF:$Fe_3O_4$ sample. It clearly reveals a superparamagnetic behavior of the $Fe_3O_4$ particles with zero remnant magnetization. Although $Fe_3O_4$ particles form clusters due to aggregation, each particle still has its own magnetization and its direction can be randomly flipped under the influence of temperature. In the absence of an external magnetic field, the average magnetization is zero. At large field, the magnetizations of all particles are aligned along the direction of the applied magnetic field. The saturation of the magnetization is obtained for an applied field of about 0.2T at 10 K and 0.7T at 300K. The saturation magnetization increases when the temperature decreases. The field cooling (FC) and zero-field cooling (ZFC) measurements (see inset of Fig. 2) allow us to deduce the information of blocking temperature and estimate the nanoparticle size.[28] The moment is measured with a small field of +5mT for ZFC and FC (cooling with +5T) conditions. From the ZFC curve, the blocking temperature can be obtained to be around 15 K. This corresponds to a 5 nm size for the $Fe_3O_4$ nanoparticles,[28] which is in a good agreement with the nominal particle size of the purchased product.

A scheme of the LSMO/PVDF:$Fe_3O_4$/Co/Au device used for the electrical measurements is schematically shown in Fig. 3(a). *I-V* measurements were performed using a two-terminal geometry by using a Keithley 2400 as a voltage source and a Keithley 6487 picoamperemeter to measure the current. To ferroelectrically polarize the PVDF barrier, we have applied different voltage pulses to the junction with a ramp of 0.1V/s and a duration of 1s. In Fig. 3(b), two distinct *I-V* curves associated with different junction resistances were obtained after applying opposite polarizing voltages. Good tunneling properties are obtained from the non-linear variation of the *I-V* characteristics, which also confirms that the $Fe_3O_4$ inclusions do not create short circuit or pinholes in the PVDF barrier. Fig. 3(c) and 3(d) show the magneto-resistance loops of the device after applying a positive and a negative polarization, respectively. While scanning the magnetic field, the magnetization of the LSMO and Co are switched separately to get either a parallel or an antiparallel magnetization configurations thus resulting in the magnetoresistance curves. The TMR is defined by the following relations: $TMR = \frac{R_{AP}-R_P}{R_P} \times 100\%$ for positive TMR, and $TMR = \frac{R_{AP}-R_P}{R_{AP}} \times 100\%$ for negative *TMR*, where $R_P$ and $R_{AP}$ are the junction resistance for parallel and antiparallel magnetizations alignment of two electrodes, respectively. The device gives TMR values of +9.2% and -11.3% when it is positively and negatively polarized, respectively. The change of the TMR sign with different PVDF polarizations could be explained by the sign change of the spin-polarization at the organic/ferromagnetic "spinterface" depending on the ferroelectric polarization of the organic barrier.[13] Another possible explanation is that some oxygen atoms could exist at Co/PVDF interface. Switching the ferroelectric polarization of PVDF shifts the Co-O antibonding state at the oxidized interface due to the screening of the polarization charges, which affects the interface transmission properties and reverses the spin-polarization at the Co/PVDF interface.[12]

In addition to the TMR effect, a clear difference between the two parallel resistances ($R_P$) for the two polarizations can be observed. The tunneling electroresistance is defined as: $TER = \frac{R_P^{Down}-R_P^{Up}}{R_P^{Up}} \times 100\%$, where $R_P^{Down}$ and $R_P^{Up}$ are the parallel resistance when the PVDF is in the "down" and "up" polarization states, respectively. The TER of the device reaches about 215%. The four resistance states associated with different magnetization and ferroelectric configurations prove the realization of the OMFTJ function with a PVDF barrier containing $Fe_3O_4$ nanoparticles. The TMR observed in PVDF:$Fe_3O_4$ OMFTJ is comparable to that of pure PVDF OMFTJ.[13] However, the TER measured in PVDF:$Fe_3O_4$ nanocomposite films is 3 times higher than that of pure PVDF OMFTJ (~75%),[13] which confirms the enhancement of the ferroelectric performance in PVDF:$Fe_3O_4$ nanocomposite based OMFTJs.

Fig. 3(e) shows the two successive loops of the parallel junction resistance $R_P$ as a function of the polarizing voltage measured at 10K. The resistance clearly shows a hysteresis loop behavior. The first loop shows a partial polarization situation with a smaller polarization switching voltage, while the second loop can be regarded as fully polarized situation, showing much higher TER (450%) compared to the first loop (293%). When the ferroelectric domains are not fully polarized with the applied electric field, a larger opposite field is necessary to switch back the polarization, thus making the loop more and more opened.[29] This validates that the observed resistance switching is due to the ferroelectricity of PVDF and excludes other mechanisms, such as the migration of oxygen vacancy in LSMO surface[30] or a reversible redox reaction of electrodes[31]. To further optimize the TER, we have fabricated several samples by varying the PVDF:$Fe_3O_4$ barrier thickness. Fig. 3(f) shows the TER as a function of the barrier thickness and the corresponding resistance for the negative polarization state. It is found that the optimized thickness is about 18 nm. Both a lower and a larger barrier thickness leads to a decreasing TER value. These results are in good agreement with our previous thickness dependent results in P(VDF-TrFE) OMFTJs.[14] We have shown that the decreasing TER value for the device with a thin barrier is mainly due to the incomplete ferroelectric polarization of the barrier.[14] For the thicker barrier, the tunneling mechanism is dominated by thermally activated hopping inside the barrier.[32] Moreover, the larger interface roughness induced by the larger thickness is also detrimental to the ferroelectric properties.[13]

Finally, we compare the temperature dependence of the TER for OMFTJs with a pure PVDF barrier and a PVDF:$Fe_3O_4$ nanocomposite barrier, as shown in Figs. 4(a) and 4(b). For both devices, we have polarized the junctions at 10K and separately measured the junction resistance with increasing temperature in both polarization states. For the device with pure PVDF (Fig. 4(a)), the resistance variations for both polarization states appear rather complicated, indicating that several mechanisms influencing the tunneling behavior are at work.[33] These mechanisms could include thermal activation of the hopping process, thermal strain in the PVDF or thermal relaxation of ferroelectric domains.[33] In the temperature range T>120K (indicated by dashed line), the resistance in the positive polarization state decreases, while it increases in the negative polarization state. The disappearing of the resistance difference for both polarization states results in a rapid drop of the TER from 80% at 120K to 7% at RT. This indicates that the thermal fluctuation of the ferroelectric domains is main mechanism to be responsible for the temperature dependence of resistance in the temperature range higher than 120K.[34] One can then estimate the energy barrier between the two polarization states to be about $k_BT$=10meV from this characteristic temperature of 120K. As a comparison, the PVDF:$Fe_3O_4$ based OMFTJ shows a much better thermal stability of the ferroelectric properties. Firstly, the characteristic temperature is shifted to 170K (indicated by the dashed line in Fig. 4(b)), indicating an increased energy barrier (corresponding to $k_BT$=14meV) between the two polarization states. Secondly, the resistance contrast decrease is much less pronounced when the temperature increases. It is noteworthy that one can still obtain 100% TER at room temperature.

The improvement of the ferroelectric properties of PVDF by adding $Fe_3O_4$ nanoparticles can be explained by the following reasons. Firstly, the interaction between the negatively charged surface of the nanoparticles and the positively charged polymer $CH_2$ groups can promote the nucleation of the polar β phase of PVDF,[23,24] which gives the main contribution to the ferroelectricity of the PVDF barrier. Secondly, the addition of $Fe_3O_4$ nanoparticles brings more position for charge accumulation at the internal interfaces. The migration and accumulation of charge carriers can result in a homogenous polarization and a high dielectric constant.[20, 35, 36, 37] Thirdly, PVDF:$Fe_3O_4$ nanocomposite exhibits multiferroic properties with significant magneto-electric (ME) coupling at PVDF/particle interfacial regions, that could result in the enhancement of the ferroelectric properties.[21,38] In addition, due to the ME coupling between the PVDF host and the $Fe_3O_4$ magnetic nanoparticles, the polarizing electric field could also alter the orientation of magnetization of $Fe_3O_4$ particles and affect the spin-dependent interface transmission probabilities.[39]

In summary, we have fabricated $La_{0.6}Sr_{0.4}MnO_3$/PVDF:$Fe_3O_4$ nanocomposite/Co OMFTJ. It is found that the ferroelectric properties of the OMFTJs can be considerably improved by adding $Fe_3O_4$ nanoparticles to the PVDF barrier. The PVDF:$Fe_3O_4$ OMFTJ shows a high TER of about 450% at 10K, which is six times larger than that of a pure PVDF based device. In addition, the higher energy barrier (14meV) between two polarization states ensures a better thermal stability of the OMFTJ with $Fe_3O_4$ nanoparticles, which can maintain a TER about 100% even at room temperature. The enhanced ferroelectricity of OMFTJ based on the PVDF:$Fe_3O_4$ nanocomposite will promote the OMFTJ for memristor and spintronics applications.

**Acknowledgements**

We acknowledge the support from the French National Research Agency (ANR) FEOrgSpin project (Grant No. ANR-18-CE24-0017) and SIZMO2D project (Grant No. ANR-19-CE24-0005-02). We also thank the PHC CAI YUANPEI 2017 (N° 38917YJ) program, ICEEL (INTER-Carnot) BlueSpinLED and ICEEL (international) SHATIPN projects. The Région Hauts-de-France and FEDER (EU) under the CPER project "Chemistry and Materials for a Sustainable Growth" is acknowledged for funding of MFP-3D microscope. A.F., A.D.C. and R.D. thank L. Maës for technical support. X.G. acknowledges the joint Ph.D scholarship from the China Scholarship Council and the Région Lorraine. X.G.L. acknowledges National Natural Science Foundation of China (NSFC, Grant Nos. 51790491) for LSMO substrates. Experiments were performed using equipment from the platform TUBE–DAUM funded by FEDER (EU), ANR, the Region Lorraine and Grand Nancy.

**Figures**

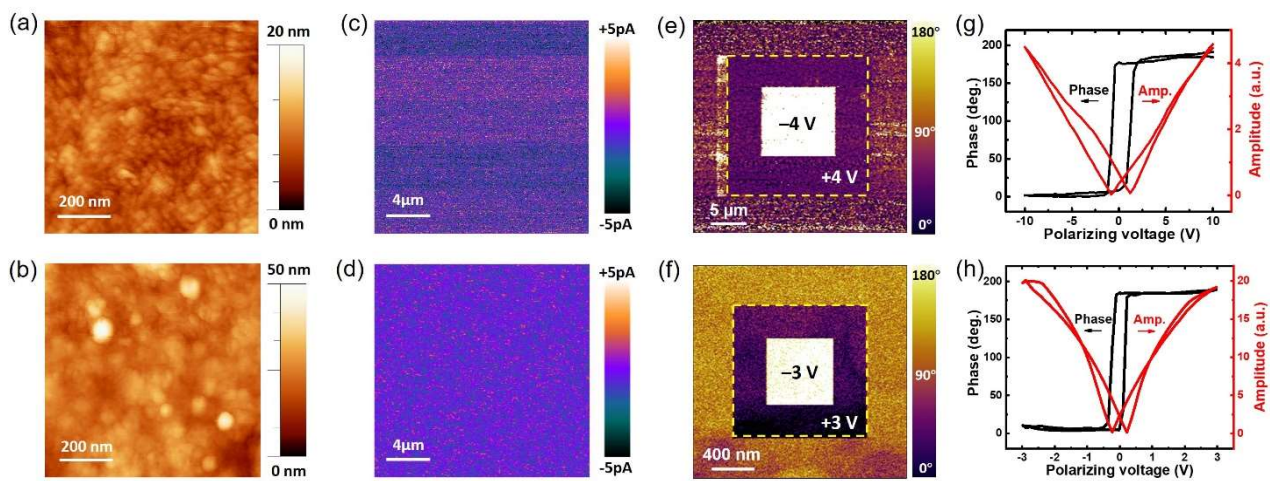

**FIG.1.** Morphology and piezoelectric characterization of pure PVDF film and PVDF:$Fe_3O_4$ nanocomposite film. (a) AFM image of a pure PVDF film. (b) AFM image of a PVDF:$Fe_3O_4$ nanocomposite film. (c-d) C-AFM scanning of barrier conductivity under an applied bias of +5V for (c) pure PVDF film and (d) PVDF: $Fe_3O_4$ nanocomposite film. (e-f) PFM phase image recorded on (e) pure PVDF film and (f) PVDF: $Fe_3O_4$ nanocomposite film. The contrasts showing the ferroelectric switching were obtained after application of a positive applied DC bias on the tip over the yellow dashed square and subsequently negative applied DC bias over the middle area. (g-h) Local PFM hysteresis phase and amplitude loops measured on (g) pure PVDF film and (h) PVDF: $Fe_3O_4$ nanocomposite film.

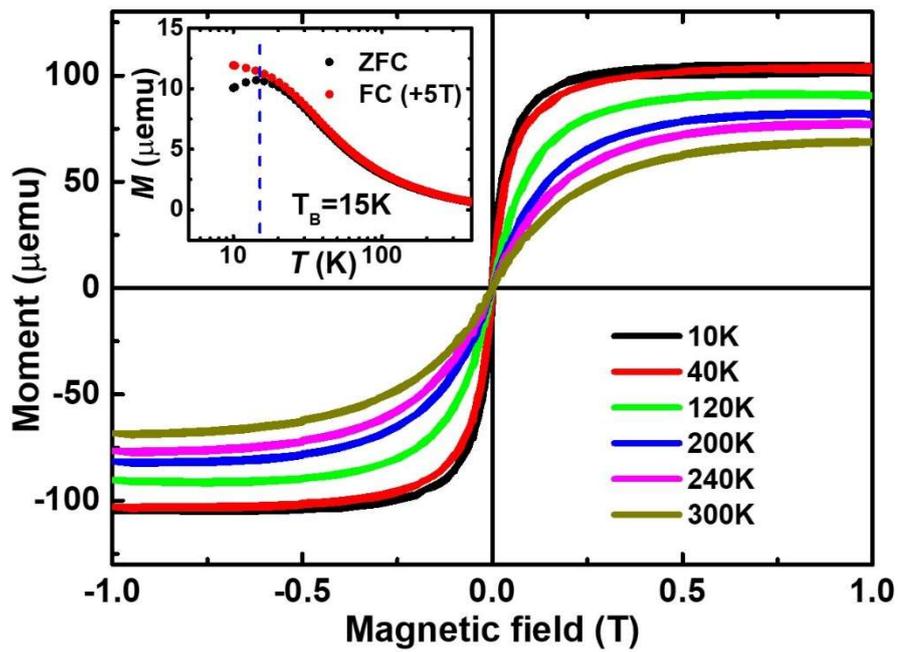

**FIG.2.** Magnetic properties of PVDF:Fe$_3$O$_4$ nanocomposite film. SQUID measurement of *M-H* curves at different temperatures for the Si/Au/PVDF:Fe$_3$O$_4$ sample. Inset: *M-T* measurements under zero field cooling (ZFC) and field cooling (FC) with +5T. The moment is measured with a +5mT field.

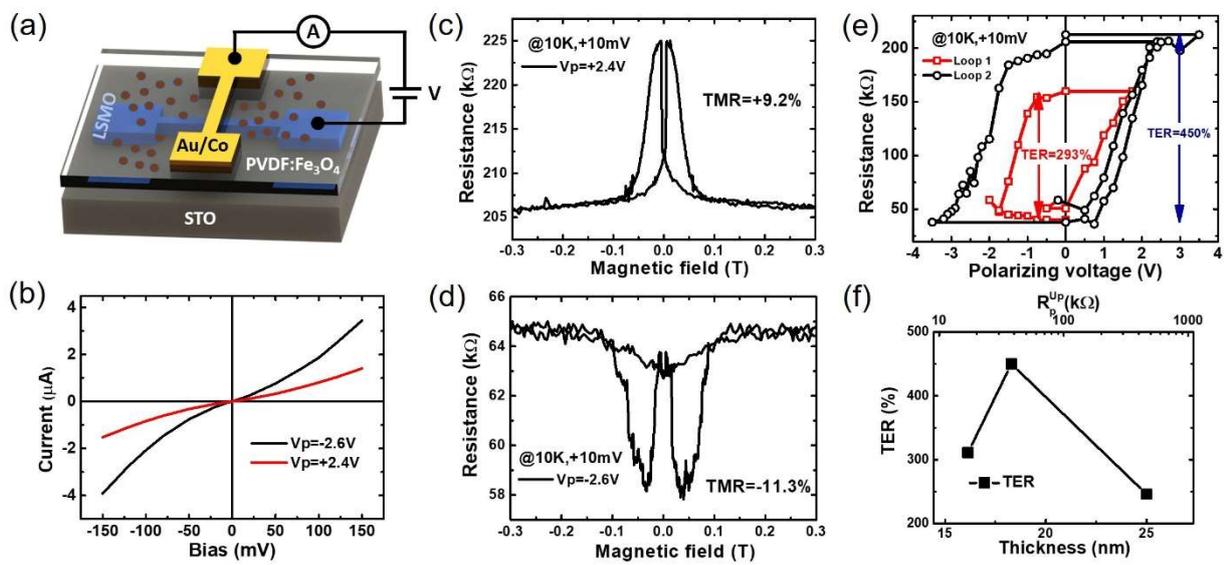

**FIG. 3.** Magneto-transport measurements on LSMO/PVDF:Fe$_3$O$_4$/Co devices. (a) Schematics of the structure and measurement setup of LSMO/PVDF:Fe$_3$O$_4$ /Co device. (b) *I-V* characteristics measured between ±0.2 V after being polarized with -2.6 V and +2.4 V. (c-d) Magnetoresistance curves measured under an applied bias of +10mV at 10K after being polarized with (c) +2.4 V and (d) -2.6 V. (e) Two successive loops for parallel resistance loop versus electric polarizing voltage. (f) TER as a function of the PVDF:Fe$_3$O$_4$ barrier thickness.

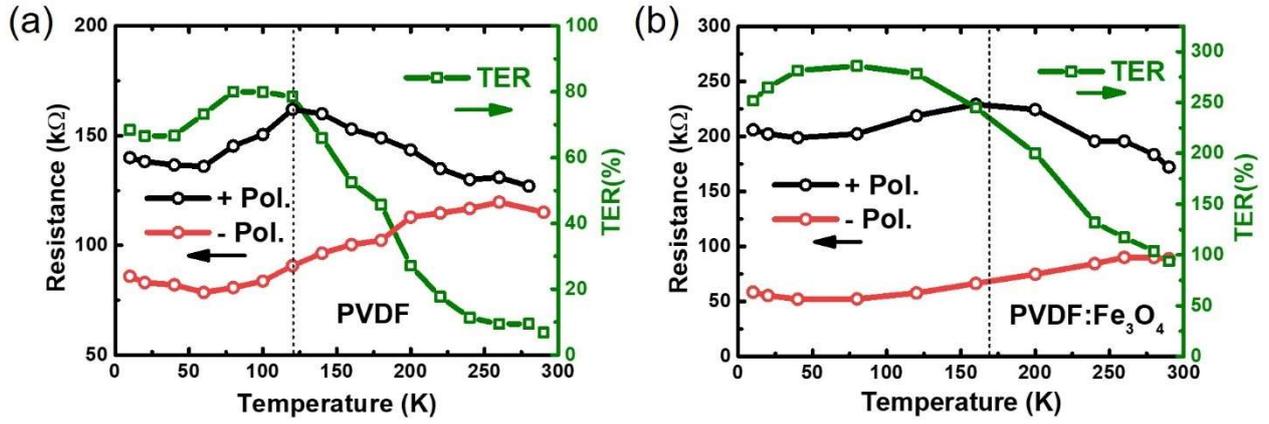

**FIG.4.** TER as a function of temperature. (a) Temperature dependence of the parallel resistance (left axis) and TER ratios (right axis) at an applied bias of +10 mV with polarizing voltage $V_P$ = +1.5 V and –1.5 V for the two PVDF polarization states. (b) Temperature dependence of parallel resistance (left axis) and TER ratios (right axis) at an applied bias of +10 mV with polarizing voltage $V_P$ = +2.4 V and –2 V for the two PVDF polarization states. The polarizing voltage is only applied at 10 K before increasing the temperature.